# Bridging the Mid-Infrared-to-Telecom Gap with Silicon Nanophotonic Spectral Translation

Xiaoping Liu[1,‡,*], Bart Kuyken[2,*], Gunther Roelkens[2], Roel Baets[2], Richard M. Osgood, Jr.[1], and William M. J. Green[3,†]

Expanding far beyond traditional applications in optical interconnects at telecommunications wavelengths, the silicon nanophotonic integrated circuit platform has recently proven its merits for working with mid-infrared (mid-IR) optical signals in the 2-8 μm range. Silicon's broadband transparency[1, 2], strong optical confinement, and potential for co-integration with CMOS electronics[3] are but a few of the many characteristics making the silicon platform ideal for development of high-performance, densely-integrated mid-IR optical systems. These systems are capable of addressing applications including industrial process and environmental monitoring[4], threat detection[5], medical diagnostics[6], and free-space communication[7]. Rapid progress has led to the demonstration of various silicon components designed for the on-chip processing of mid-IR signals, including waveguides[8-10], vertical grating couplers[11], microcavities[12, 13], and electrooptic modulators[14]. Even so, a notable obstacle to the continued advancement of chip-scale systems is imposed by the narrow-bandgap semiconductors, such as InSb and HgCdTe, traditionally used to convert mid-IR photons to electrical currents. The cryogenic or multi-stage thermo-electric cooling required to suppress dark current noise[15], exponentially dependent upon the ratio $E_g/kT$, can limit the development of small, low-power, and low-cost integrated optical systems for the mid-IR. However, if the mid-IR optical signal could be spectrally translated to shorter wavelengths, for example within the near-infrared telecom band, photodetectors using wider bandgap semiconductors such as InGaAs or Ge could be used to eliminate prohibitive cooling requirements. Moreover, telecom band detectors typically perform with higher detectivity and faster response times when compared with their mid-IR counterparts[15]. Spectral translators employing sum or difference frequency generation in nonlinear crystals, including $LiNbO_3$[16-18] and KTP[19], have been studied. However, such systems can be impeded by low conversion efficiencies, their significant size, and limited integrability of their component parts. Here we address these challenges with a silicon-integrated approach to spectral translation, by employing efficient four-wave mixing (FWM) and large optical parametric gain in silicon nanophotonic wires[20-22]. Using an optical pump near silicon's two-photon absorption (TPA) threshold[23], we excite nanophotonic wires uniquely engineered to use higher-order waveguide dispersion to facilitate spectral translation of a mid-IR input signal at 2440 nm to the telecom band at 1620 nm, across a span of 62 THz. The converted signal simultaneously experiences a translation gain of more than 19 dB, an efficiency enhancement which can dramatically improve the detection sensitivity for weak mid-IR signals. Moreover, this single silicon device also performs as a transmitter, by converting telecom band signals to the mid-IR with a translation gain of 8.0 dB. Finally, an 8.4 dB optical parametric amplification of telecom band signals is demonstrated when using a mid-IR pump, reinforcing the wide-ranging technological role silicon nanophotonic wires can serve within both the mid-IR and telecom bands.

[1]Department of Electrical Engineering, Columbia University, 1300 S. W. Mudd Building, 500 W. 120th Street, New York, New York 10027, [2]Photonics Research Group, Department of Information Technology, Ghent University–Imec, Ghent B-9000, Belgium, [3]IBM T. J. Watson Research Center, Yorktown Heights, NY 10598, USA, ‡Current address: OFS Labs, 19 Schoolhouse Road, Somerset, NJ 08873, USA. *These authors contributed equally to this work. †wgreen@us.ibm.com

Mid-IR to telecom band spectral translation in silicon wires can be accomplished using efficient FWM with discrete band phase-matching[24, 25]. In this process, the pump is placed away from the zero dispersion wavelength, and higher-order waveguide dispersion is used to phase-match a discrete pair of bands at spectrally distant frequencies, located symmetrically on either side of the pump. Discrete band phase-matching can be achieved in a waveguide with anomalous 2$^{nd}$-order dispersion ($\beta_2 < 0$) and small positive 4$^{th}$-order dispersion ($\beta_4 > 0$) (see Supplementary Figure 1), conditions which are engineered[21] through manipulating the cross-sectional dimensions and cladding materials of the silicon nanophotonic wire. Figure 1a shows an optical microscope image of the 2 cm long silicon nanophotonic wire used here for spectral translation, as fabricated on a 200 mm silicon-on-insulator (SOI) wafer at *imec*, through the multi-project-wafer service ePIXfab (www.ePIXfab.eu). The entire length of the wire is coiled into a compact spiral, occupying an on-chip footprint of only 625 μm by 340 μm. The wire has cross-sectional dimensions of $w$ = 900 nm by $h$ = 220 nm, as shown in Fig. 1b. The cladding consists of air above and a 2 μm buried oxide (SiO$_2$) below the silicon core. Numerical simulations indicate that the dispersion conditions $\beta_2 < 0$ and $\beta_4 > 0$ are satisfied over the spectral range from 1810-2410 nm (see Supplementary Figure 2). Over a similar range, a large effective nonlinearity parameter of $\gamma \sim$ 130 (W·m)$^{-1}$ and a low propagation loss of 2.6 dB/cm also serve to facilitate highly efficient FWM for this compact device.

The nonlinear mixing and spectral translation characteristics of the silicon nanophotonic wire are illustrated in Figs. 1c-d. Figure 1c shows the recorded output spectrum when the wire is excited by a pump pulse-train at 1946 nm having a peak coupled input power of 37.3 W (for experimental details see Methods). While this pump wavelength is not yet beyond silicon's TPA threshold of 2200 nm[23], the TPA coefficient is nevertheless a factor of 2-3x lower than that at 1550 nm, resulting in small nonlinear loss and efficient FWM. For example, in the absence of any probe signal (Signal OFF), the pump transmission spectrum already exhibits clear signatures of the desired phase-matched FWM processes. Specifically, strong broadband modulation instability (MI) peaks[22] appear adjacent to the pump at 1810 nm and 2090 nm. Moreover, two additional discrete MI bands with much larger detuning from the pump appear at 1620 nm and 2440 nm, and serve as direct evidence of higher-order phase-matching. The absolute power of the MI band at 2440 nm appears lower than that of the MI band at 1620 nm, due to a 1.8 dB asymmetry expected from the Manley-Rowe power division relations, as well as from ~3-4 dB larger losses in the output fibre optical collection path at longer wavelengths.

The visibility of the MI bands, associated with the parametric amplification of background noise, suggests that the pumped silicon nanophotonic wire should exhibit significant parametric gain as well as a large conversion efficiency when probed by input signals at these wavelengths. Figure 1d illustrates the output spectrum in one such case, when the long-wavelength discrete MI band is probed (Signal ON) by a low-power ($P_{sig}$ < 35 µW) continuous-wave mid-IR signal at 2440 nm. When the signal is tuned into resonance with this spectral band, it experiences strong parametric amplification through degenerate FWM (evidenced by the appearance of the large spectral pedestal), and is simultaneously up-converted to a prominent telecom band idler at 1620 nm. This large spectral translation over more than 62 THz illustrates that the higher-order dispersion design methodology applied here may be used to efficiently convert optical information on a mid-IR carrier into the telecom band, where it can be detected and processed using un-cooled, high-speed, high-sensitivity III-V and group-IV semiconductor detector technologies.

By recording transmission spectra for a range of signal wavelengths near 2440 nm and 1620 nm, the wavelength dependence of spectral translation efficiency and parametric gain within the discrete phase-matching bands can be determined (see Methods). Figure 2a illustrates that for mid-IR input signals, the silicon wire device attains optical transparency (on-chip gain exceeding 0 dB) across a bandwidth of 150 nm near the signal and 45 nm near the idler. The data also demonstrates that spectral translation of mid-IR signals to the telecom band idler near 1620 nm takes place with a peak conversion gain of 19.5 dB. Therefore, not only could such a spectral translator facilitate detection of mid-IR signals without cumbersome cooling requirements, the associated optical gain could also dramatically improve the sensitivity of such a receiver system, particularly for weak mid-IR input levels[26]. At the same time, the mid-IR input signal experiences a peak on-chip parametric amplification of 18.8 dB.

Figure 2b shows the result of a similar set of experiments, in which a telecom band input signal is tuned across the MI peak near 1620 nm. In this case, the on-chip transparency bandwidth is approximately 20 nm for the signal and 40 nm for the idler. The telecom band signal is spectrally translated to the mid-IR with a peak gain of 8.0 dB, a process which can be applied to generating and transmitting arbitrary mid-IR signals using commercially-available telecom components[27, 28]. In addition to performing the spectral translation function, Fig. 2b illustrates that the telecom band input signal is simultaneously amplified by 8.4 dB. The demonstration of a silicon wire amplifier which utilizes a mid-IR pump to provide substantial parametric gain to a telecom band signal is of particular technological significance, as such an amplifier could find applications within the CMOS-integrated silicon nanophotonic platforms currently being developed for high-speed optical interconnect systems[3].

The spectral translation of the telecom band signal to the mid-IR shown in Fig. 2b occurs with an approximately 11 dB gain reduction when compared with the reversed scenario illustrated in Fig. 2a. As the energetic combination of a 1620 nm signal photon with a pump photon lies significantly above silicon's bandgap, the observed asymmetry is expected due to the effects of non-degenerate TPA[23, 24] in the silicon wire. Non-degenerate TPA produces larger attenuation of an input signal near 1620 nm as compared to one near 2440 nm, when each is combined with the strong pump at 1946 nm. Therefore, larger gain values could be expected for the telecom band signal if the pump wavelength were increased.

The phase-matched signal and idler wavelengths linked by the discrete FWM spectral translation process are dependent upon the magnitudes of $\beta_2$ and $\beta_4$, and can therefore be tuned by selection of the wavelength at which the dispersive silicon nanophotonic wire is pumped. The spectral separation between bands is expected to increase for large values of $|\beta_2|$ and small values of $|\beta_4|$. Figure 3a plots the spectral locations at which the discrete signal and idler MI bands appear as a function of the pump wavelength. The figure illustrates that the bands are maximally separated by 865 nm when the wire is pumped at 1998 nm. Conversely, the bands are observed to move closer together as the pump approaches the two zero-GVD points located at 1810 nm and 2410 nm.

An optimization of the silicon nanophotonic wire dispersion design, focusing upon increasing $|\beta_2|$ while simultaneously decreasing $|\beta_4|$, can facilitate translation of even longer wavelength mid-IR signals into the telecom band. For example, Fig. 3b plots numerical calculations of the phase-matched discrete signal and idler wavelengths of a design tailored for translating a range of mid-IR signals from 3000-3550 nm into the L-band (see Methods). This spectrum is targeted for its overlap with a mid-IR low-loss window in $SiO_2$[1]. This design consists of a silicon wire with a thickness of 300 nm and a width in the range of 700-900 nm, completely surrounded by an oxide cladding (see Supplementary Figure 4). In the specific case of a wire with $w = 900$ nm and $h = 300$ nm, Fig. 3b illustrates that an input signal at 3550 nm could be spectrally translated to an L-band idler at 1590 nm (and vice versa), using a pump wavelength of 2200 nm. This corresponds to a span of 104 THz, more than an octave in optical frequency.

Moreover, while a pulsed pump with a high peak power is used in the present demonstration, a practical spectral translation system will require a continuous-wave (c.w.) pump, in order to eliminate any requirement for synchronization of the input mid-IR signals to the pump repetition period. Because the oxide-clad wire's dispersion characteristics were designed with the intent of operating the pump at or beyond silicon's TPA threshold at 2200 nm, c.w. parametric gain is made possible by avoiding the deleterious effects of nonlinear loss and TPA-induced free-carrier absorption. Detailed calculations described in the Supplementary Methods and plotted in Supplementary Figure 5 show that c.w. spectral translation of a 3550 nm input signal to 1590 nm can be accompanied by a translation gain as large as 22 dB, for a moderate pump power of 300 mW at 2200 nm.

In conclusion, we have shown that judicious engineering of FWM processes in silicon nanophotonic wires can facilitate amplified bi-directional spectral translation of signals between the mid-IR and the telecom band, across a 62 THz span in optical frequency. Telecom band detection of translated mid-IR signals can eliminate the burdensome cooling requirements traditionally associated with mid-IR photodetectors, and can be performed by on-chip photodetectors integrated via heterogeneous[29] or monolithic[30] techniques. Moreover, these spectral translation devices can be integrated with additional mid-IR and/or telecom band silicon nanophotonic components such as modulators, wavelength (de-)multiplexers, and switches, which together have the potential to produce flexible, chip-scale optical systems for mid-IR applications.

**Methods:**

**Four-wave mixing experimental configuration:**

In our experiments, the FWM pump is a picosecond pulse train (FWHM ~ 2 ps, repetition rate = 76 MHz) from a tunable optical parametric oscillator (mode-locked Ti:sapphire-pumped Coherent Mira-OPO). The pump spectrum has a signal-to-noise ratio larger than 70 dB over the wavelength range from 1600 nm to 2500 nm (see Supplementary Figure 3). The c.w. probe signals are obtained either from a tunable mid-IR laser (IPG Photonics SFTL $Cr^{2+}$:ZnSe polycrystal with erbium-fibre laser pump source) or a tunable telecom laser (Agilent 81640A). Pump and probe are coupled into two separate single-mode optical fibres, aligned individually with polarization controllers to excite the fundamental quasi-TE mode in the silicon nanophotonic wire, and then multiplexed with a 90/10 fused fibre directional coupler. Coupling into/out of the 2 cm-long silicon nanophotonic wire is accomplished via cleaved facet edge coupling with lensed tapered fibres (coupling losses = 10 +/- 1 dB/facet, across the entire spectral range utilized). The spectral content of the transmitted light is analyzed with 1 nm spectral resolution, using a mid-IR optical spectrum analyzer (Yokogawa AQ6375).

**Extraction of spectral translation efficiency and signal gain:**

The peak power of the spectrally translated idler pulse at the output of the silicon nanophotonic wire, $P_{idler\_out}$, is derived from the measured FWM spectra, according to $P_{idler\_out} = F\left(\int P_{idler\_avg}(\lambda)d\lambda\right)$. In order to convert the time-averaged idler power $P_{idler\_avg}$ measured by the OSA into peak power, the spectrally integrated power is weighted by the duty cycle factor $F = 1/(76\text{ MHz} \cdot 2\text{ ps})$ to account for the pulsed nature of the experiment. A similar procedure is applied to calculate the signal output power $P_{signal\_out}$. A 2 nm wide band-stop filter is first numerically applied to the time-averaged signal spectrum, in order to reject the power remaining in the narrowband c.w. tone. The peak signal power is then computed according to $P_{signal\_out} = F\left(\int P_{signal\_filtered\_avg}(\lambda)d\lambda\right)$. Finally, to find the c.w. signal power at the waveguide input $P_{signal\_in}$, the output c.w. signal power is measured (with the pump off) and corrected to account for total propagation losses of $\alpha$ dB incurred through the 2 cm long device, $P_{signal\_in} = 10^{\alpha/10}\left(\int P_{signal\_out\_pump\_off}(\lambda)d\lambda\right)$. The on-chip idler spectral translation gain $\eta$ is then defined as the ratio of peak idler power and input c.w. signal power, $\eta = P_{idler\_out}/P_{signal\_in}$. Accordingly, the on-chip parametric signal gain is given by $G = P_{signal\_out}/P_{signal\_in}$. The error bars in the on-chip parametric gain data are calculated to reflect the uncertainty in the total propagation loss $\alpha$, as well as the contribution of the MI noise background accumulated when integrating the signal/idler power at the waveguide output. Additional detail on the estimation of error bars is included in the Supplementary Methods.

**Numerical calculations of signal and idler discrete band tuning versus pump wavelength:**

The linear phase-mismatch $\Delta k_l$ is characterized by $\Delta k_l = \beta_s + \beta_i - 2\beta_p$, where $\beta_s$, $\beta_i$, and $\beta_p$ are the modal propagation constants for signal, idler, and pump respectively. These propagation constants and their spectral dispersion are obtained through numerical simulations, using a commercial finite-element eigen-mode solver (RSoft FemSim). Solving the phase-matching equation $\Delta k_l + 2\gamma P = 0$ ($P = 300$ mW is assumed for the calculation in Fig. 3b) with signal and idler frequencies constrained by the pump detuning $\Delta\omega = |\omega_p - \omega_s| = |\omega_p - \omega_i|$ will generally yield two solutions for $\Delta\omega$. The smaller-valued solution corresponds to broadband phase-matched signal-idler regions appearing immediately adjacent to the pump, and is ignored for the purposes of spectral translation in this work. The larger-valued of the two solutions describes the discrete signal and idler bands appearing at large pump detuning. The curves of phase-matched signal-idler pairs shown in Fig. 3b are found by repeating this calculation for a range of pump frequencies, and converting to detuning in units of wavelength.

**End Notes:**

**Acknowledgements**

The authors gratefully acknowledge the staff at *imec* (Leuven, Belgium) where the silicon nanophotonic waveguide devices were fabricated. They would also like to thank Y. A. Vlasov (IBM Research) for many helpful and motivating discussions. B.K. acknowledges the Flemish Research Foundation (FWO) for a doctoral fellowship. This work was partially funded under the Methusalem "Smart Photonic Chips," FP7-ERC-INSPECTRA, and FP7-ERC-MIRACLE programs.

**Author contributions**

X.L. (xl2165@columbia.edu) and B.K. (bart.kuyken@intec.ugent.be) performed the numerical dispersion and phase-matching design calculations. B.K., G.R., and R.B. supervised the waveguide device fabrication process. X.L. and B.K. performed the wavelength translation experiments with guidance and supervision from W.M.J.G. All authors contributed to the data analysis and writing of the manuscript.

**Author information**

The authors declare no competing financial interests. Correspondence and requests for materials should be addressed to W.M.J.G. (wgreen@us.ibm.com).

**Figure legends:**

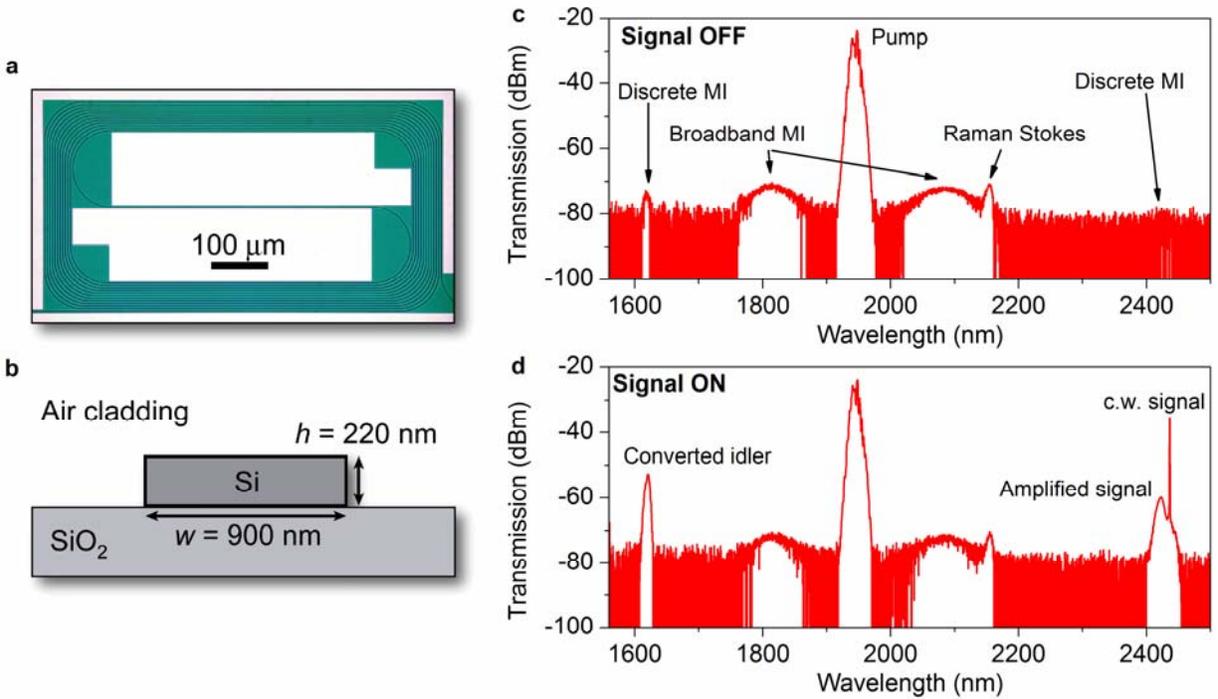

**Figure 1 | Structural design and transmission characteristics of the silicon nanophotonic wire spectral translation device. a**, Optical microscope image of the spiral-coiled silicon wire. The wire has a total length of 2 cm, and occupies an on-chip footprint of only 625 μm x 340 μm. Bends with a conservative 60 μm radius are used. **b**, Cross-sectional schematic, illustrating a silicon core with a width of $w$ = 900 nm and a height of $h$ = 220 nm, which lies upon a 2 μm thick $SiO_2$ buried oxide layer. The silicon core is air-clad from above. **c**, Output transmission spectrum with pump operating at λ = 1946 nm and input signal OFF. The observed modulation instability spectrum generated by amplification of background noise serves as a marker of the spectral bands in which phase-matching conditions are met. The location of the broadband MI peaks adjacent to the pump at 1810 nm and 2090 nm is primarily determined by $β_2$, while the discrete MI bands at 1620 nm and 2440 nm occur as a result of higher-order phase-matching

dictated by the values of both $\beta_2$ and $\beta_4$. A Raman Stokes peak is also observed at 2155 nm. **d**, Transmission spectrum with input signal ON. A c.w. mid-IR signal is tuned to coincide with the discrete MI band at 2440 nm. Parametric amplification of the signal occurs with simultaneous spectral translation across 62 THz, to an idler at 1620 nm.

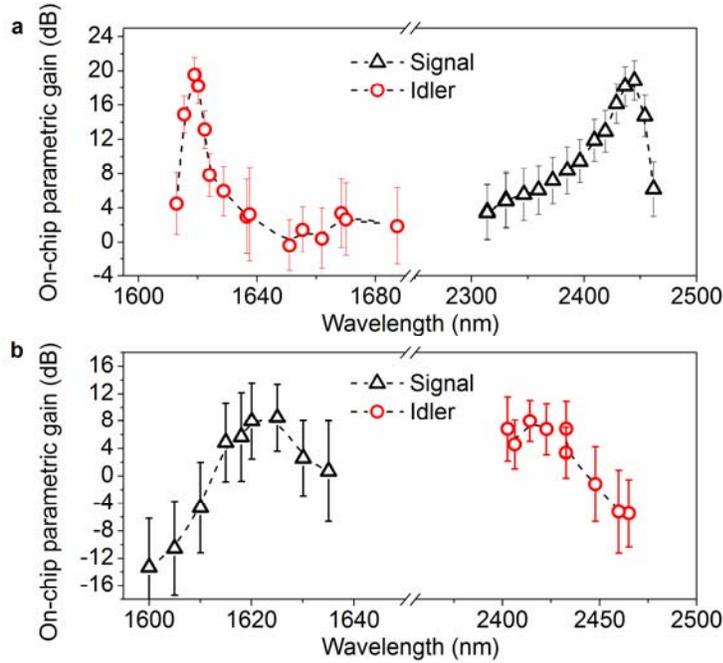

**Figure 2 | Wavelength-resolved on-chip spectral translation efficiency and parametric signal gain. a**, Injection of a mid-IR input signal ($P_{sig} < 35$ μW), with translation to a telecom band output idler. The peak on-chip translation efficiency is 19.5 dB, while the signal gain is 18.8 dB. The transparency bandwidth exceeds 45 nm near the idler, and 150 nm near the signal. **b**, Reversed scenario, with injection of a telecom band input signal ($P_{sig} < 50$ μW) and translation to a mid-IR output idler. The peak on-chip translation efficiency is 8.0 dB, and the signal gain is 8.4 dB. Transparency is reached over a bandwidth of 40 nm near the idler and 20 nm near the signal. In all of the above measurements, the silicon nanophotonic wire is pumped at 1946 nm with a peak power of 37.3 W. The small shift in the spectral position of the mid-IR gain peak between Fig. 2a and Fig. 2b (2440 nm versus 2420 nm, respectively) occurs as a result of pump wavelength drift. The dashed curves are included as a guide to the eye. Estimation of error bars is described in Supplementary Methods.

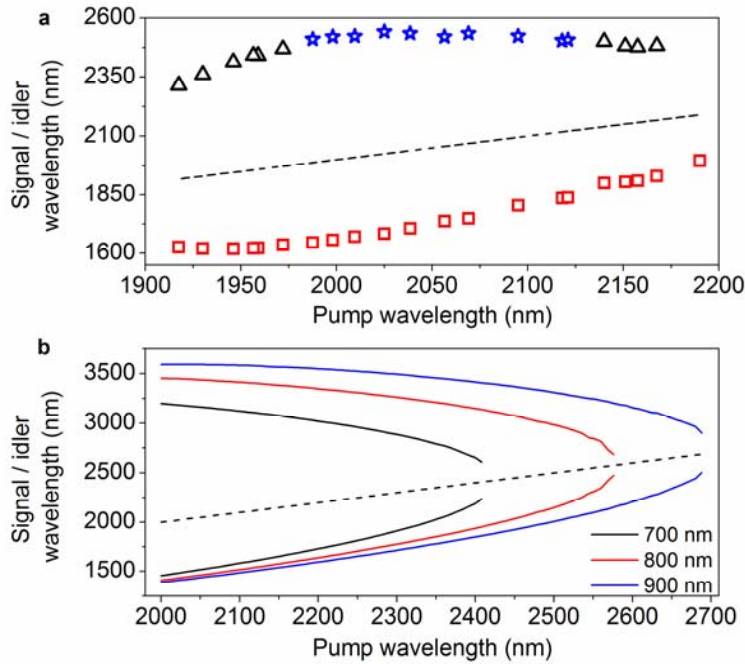

**Figure 3 | Phase-matched signal and idler wavelengths linked by the silicon nanophotonic spectral translation process. a**, The symbols mark the spectral locations of the discrete MI bands as a function of pump wavelength, for the experimentally demonstrated silicon wire having $w$ = 900 nm, $h$ = 220 nm. The MI bands indicated by blue stars were not measured directly, as they were located beyond the 2500 nm maximum wavelength limit of the spectrum analyzer used. The positions of these bands were therefore inferred from energy conservation. **b**, Design calculations describing the phase-matched discrete band locations versus pump wavelength, for the fundamental quasi-TE mode of an $SiO_2$-clad silicon wire with $h$ = 300 nm and widths $w$ = 700 nm, 800 nm, and 900 nm. The wires are tailored for spectral translation across more than an octave in optical frequency, between the 3000-3550 nm mid-IR range and the L-band. The calculations assume c.w. pumping with 300 mW pump power. In both panels, the dashed line marks the pump wavelength.

**Supplementary Methods:**

**Higher-order phase-matching for spectral translation:**

To achieve high efficiency in the FWM process, a phase-matching condition must be satisfied[1]. In the case of degenerate FWM ($2\omega_p = \omega_s + \omega_i$), the phase matching condition becomes

$$\Delta k = \Delta k_{nl} + \Delta k_l = 0 \Rightarrow 2\gamma_p P_p = -\Delta k_l \qquad (1)$$

The linear propagation constant/phase-mismatch between two pump photons ($2\beta^p$), one signal photon ($\beta^s$), and one idler photon ($\beta^i$) must be offset by the pump-induced nonlinear phase shift $2\gamma_p P_p$ caused by self-phase modulation and cross-phase modulation. Here $\gamma_p$ is the waveguide effective nonlinearity parameter, and $P_p$ is the peak pump power in the silicon nanophotonic photonic wire. For a small frequency detuning from the pump, the linear phase mismatch can be approximated using the second-order waveguide dispersion ($\beta_2^p$) at the pump frequency, such that $\Delta k_l = \beta^s + \beta^i - 2\beta^p \approx \beta_2^p (\Delta\omega)^2$ where $\Delta\omega = |\omega_p - \omega_s| = |\omega_p - \omega_i|$ is the frequency detuning between the pump and the signal/idler waves. In this case, anomalous dispersion with negative $\beta_2^p$ can lead to a phase-matched solution to Eq. (1). Indeed, parametric gain has been achieved in silicon wires within a spectral region adjacent to the pump, by excitation within the anomalous dispersion regime[2, 3].

However, when considering large values of detuning $\Delta\omega$, for example, in the case of spectral translation between a mid-IR signal and a telecom band idler, it becomes necessary to take higher-order dispersion terms into account. When dispersion up to the 4$^{th}$-order $\beta_4^p$ is retained, Eq. (1) becomes

$$2\gamma_p P_p = -\beta_2^p (\Delta\omega)^2 - \beta_4^p (\Delta\omega)^4 /12 \qquad (2)$$

Supplementary Fig. 1 depicts the graphical solutions to Eq. (2) for various combinations of the sign of the dispersion coefficients $\beta_2^p$ and $\beta_4^p$. As this figure is intended to illustrate the phase-matching conditions in general, specific numerical values are not considered. Solutions appear where the negative linear phase mismatch -$\Delta k_l$ represented by the red curves intersects the constant value $2\gamma_p P_p$. First, Supplementary Fig. 1a shows that when both $\beta_2^p$ and $\beta_4^p$ are positive, there is no solution to Eq. (2). However, there does exist one phase-matching point within the normal dispersion regime as shown in Supplementary Fig. 1b, where $\beta_2^p$ remains positive while $\beta_4^p$ is negative. This condition has been used to achieve parametric amplification in fiber OPAs[4] and OPOs[5] which are pumped in the normal dispersion regime. The case where both $\beta_2^p$ and $\beta_4^p$ exhibit negative values is shown in Supplementary Fig. 1c, where again there is only one phase-matching point.

Supplementary Fig. 1d shows the case relevant to the current work, where $\beta_2^p$ is negative and $\beta_4^p$ is positive. Depending upon the relative magnitudes of $\beta_2^p$ and $\beta_4^p$, a variety of scenarios are possible. In some cases, as is shown by the example of the dotted red curve, phase-matching is not possible due to the fact that the 4$^{th}$-order linear phase-mismatch term $-\beta_4^p (\Delta\omega)^4 /12$ dominates the contribution from 2$^{nd}$-order dispersion $-\beta_2^p (\Delta\omega)^2$, preventing the total linear phase-mismatch from reaching a value near $2\gamma_p P_p$. However, if the 4$^{th}$-order dispersion is somewhat smaller, the contributions of the two dispersion orders can balance each other in the vicinity of $-\Delta k_l = 2\gamma_p P_p$, as shown by the dashed red curve. In this case, there is a broad frequency-detuning range (indicated by the dotted blue arrow) where the peak of the linear phase-mismatch curve lies entirely within the shaded grey region $\delta k = 2\gamma_p P_p \pm 2\pi/L$. This shaded area depicts the phase detuning tolerance $\pm 2\pi/L$ originating from the finite length $L$ of the silicon nanophotonic wire, within which efficient FWM can be achieved[1]. This condition can enable very broadband parametric processes, extending over a continuous spectrum on either side of the pump. Finally, if the contribution from $\beta_2^p$ to the linear phase-mismatch dominates the contribution from $\beta_4^p$, the peak linear phase-mismatch will reach its maximum well above $2\gamma_p P_p$, as shown by the solid red curve. This scenario gives rise to a discrete and narrow phase-matching band at large detuning from the pump[6, 7], which is indicated by the arrow labeled "2nd PMB" in Supplementary Fig. 1d. This phase-matched solution is responsible for the appearance of the discrete MI bands in our experiments, and is utilized here for the purpose of spectral translation with simultaneous amplification. Equation (2) and Supplementary Fig. 1d illustrate that in order to increase the frequency span for spectral translation, it is desirable to design the silicon wire to have large values of $|\beta_2^p|$ and small values of $|\beta_4^p|$. A broader phase-matching band labeled "1st PMB" in Supplementary Fig. 1d is also observed experimentally, and is highlighted by the broadband MI regions adjacent to the pump.

**Calculation of silicon nanophotonic wire dispersion:**

The wavelength-dependent modal effective index $\bar{n}(\lambda)$ of the silicon wire's fundamental quasi-TE mode is calculated using a fully-vectorial commercial finite-element mode solver (RSoft FemSim), with 1 nm spectral resolution[8, 9]. The $\bar{n}(\lambda)$ data is fitted with a 9th-order polynomial, which is subsequently converted to a function of angular-frequency, $\bar{n}(\omega) = \sum_{i=0}^{9} a_i \omega^i$, where $a_i$ is the $i^{th}$ coefficient for the polynomial. The second- and fourth-order dispersion coefficients $\beta_2$ and $\beta_4$ are calculated via the expression $\beta_n = d^n \beta_0 / d\omega^n$, where $\beta_0 = \omega \bar{n}(\omega)/c$ is the modal propagation constant, and $c$ is the speed of light in vacuum. As shown in Supplementary Fig. 2, the engineered waveguide dimensions produce anomalous dispersion conditions ($2^{nd}$-order dispersion $\beta_2 < 0$) between the two zero-dispersion wavelengths of 1810 nm and 2410 nm. Furthermore, this silicon nanophotonic wire has small positive $4^{th}$-order dispersion ($\beta_4 > 0$) within the same wavelength range. Therefore, the opposing signs of $\beta_2$ and $\beta_4$ should permit discrete phase-matching when this silicon nanophotonic wire is pumped within the wavelength range between 1810 nm and 2410 nm.

**On-chip parametric gain error bar estimation:**

The procedure by which the error bars shown in Fig. 2 are estimated is described below, using the on-chip idler spectral translation gain as an example. A similar procedure was followed for the parametric signal gain.

As previously defined in Methods, the on-chip idler conversion gain is given by $\eta = P_{idler\_out} / P_{signal\_in} = F \left( \int P_{idler\_avg}(\lambda) d\lambda \right) / \left[ 10^{\alpha/10} \left( \int P_{signal\_out\_pump\_off}(\lambda) d\lambda \right) \right]$. An error analysis of this expression shows that the uncertainty in $\eta$ consists of two uncorrelated sources, a) the uncertainty in the total propagation loss through the 2 cm long waveguide, defined as $\delta_1$, and b) the uncertainty in calculating the integrated average power using the spectra recorded by the OSA, defined as $\delta_2$. The duty-cycle factor $F$ is free of uncertainty, as the pump pulse repetition rate is measured accurately by a frequency counter and an oscilloscope, while the pulse duration is measured precisely by autocorrelation. Since we report the data and error bars in decibel (dB) units, the uncertainty in the ratio $\eta$ is given by the sum of $\delta_1$ and $\delta_2$, both in dB units.

Over the spectral range in which the input signals are swept (2314 nm < $\lambda$ < 2462 nm in the mid-IR, and 1600 nm < $\lambda$ < 1635 nm in the telecom band), our cutback waveguide propagation loss measurements show that the average propagation loss is 2.63 dB/cm, with an uncertainty of +/- 0.5 dB/cm. These numbers translate into a total average propagation loss of $\alpha$ = 5.26 +/- 1 dB through the 2 cm long waveguide (i.e. $\delta_1$ = +/- 1 dB). Our estimate of the propagation loss uncertainty is deliberately conservative (i.e. larger rather than smaller), specifically in order to ensure that we do not overestimate the value of parametric gain, and that we do not underestimate the magnitude of the error bars in Fig. 2.

To determine the error due to noise $\delta_2$, a reference OSA spectrum with only the pump turned on is recorded (similar to Fig. 1c). The dominant noise contribution in the case studied here is the spontaneous modulation instability (MI). The MI noise power contributing to the idler/signal within the same spectral bandwidth over which the integrated terms in $\eta$ are evaluated, is extracted from this reference spectrum. The idler error (in dB units) is then given by $\delta_2 = 10\log[(P_{noise} + P_{idler\_out})/P_{idler\_out}]$. The signal error from OSA noise takes a similar form.

The total on-chip parametric gain error bar $\varepsilon$ is thus given by $\varepsilon = \pm(\delta_1 + \delta_2)$. In the vicinity of the peak signal and idler gain in Fig. 2, this error is generally smallest, since the signal/idler power is much higher than the MI noise background. The relative contribution of MI noise becomes larger for wavelengths far from the gain peaks, at which the amplified signal or idler can be of similar magnitude to the noise floor.

**Calculation of continuous-wave spectral translation efficiency and signal gain:**

The oxide-clad silicon nanophotonic wire shown in Supplementary Fig. 4a with $w = 900$ nm and $h = 300$ nm is used to illustrate the case of discrete band spectral translation with a continuous-wave pump at 2200 nm, a mid-IR input signal at 3550 nm, and a telecom band idler at 1590 nm. Since the three waves span a spectrum of more than an octave, the assumption that the effective nonlinearity parameter γ is equal for all waves is no longer valid. The complex $\gamma_{ijkl}$ parameters entering the degenerate FWM process are listed in Supplementary Table 1, where the subscripts p, s, and i, denote pump, signal, and idler respectively. These are calculated by a weighted integral of silicon's bulk nonlinear susceptibility $\chi^{(3)}(-\omega_i;\omega_j,-\omega_k,\omega_l)$ over the electric fields $E_i$, $E_j$, $E_k$ and $E_l$ of the nanophotonic wire's fundamental quasi-TE modes[10, 11]. The nonlinear susceptibility $\chi^{(3)}(-\omega_i;\omega_j,-\omega_k,\omega_l)$ is calculated as the susceptibility at the average[6] angular frequency $\omega = (\omega_i + \omega_j + \omega_k + \omega_l)/4$. Note that the parameter $\gamma_{iiii}$ has a significant imaginary component due to large degenerate TPA seen by the idler at 1590 nm. The parameter $\gamma_{iipp}$ also has a non-zero imaginary component as a result of non-degenerate TPA between the pump and the idler.

**Supplementary Table 1 | Computed values of the effective nonlinearity parameter for degenerate FWM with a pump at 2200 nm, an input signal at 3550 nm, and an idler at 1590 nm. The subscripts p, s, and i are used to denote pump, signal, and idler respectively.**

| Effective nonlinearity parameter | Value (W·m)$^{-1}$ |
|---|---|
| $\gamma_{ssss}$ | 24.73 |
| $\gamma_{pppp}$ | 139.62 |
| $\gamma_{iiii}$ | 161.45+42.64i |
| $\gamma_{iipp}$ | 287.99+18.40i |
| $\gamma_{sspp}$ | 62.83 |
| $\gamma_{spip}$ | 43.13 |
| $\gamma_{ipsp}$ | 140.29 |

The coupled wave equations governing the FWM process are then solved[6], taking into account linear propagation loss, degenerate and non-degenerate TPA, as well as absorption from TPA-generated free carriers. Free-carriers are assumed to have a lifetime of 1 ns within the silicon nanophotonic wire[12, 13]. The signal at 3550 nm is taken to have a power of 1 μW at the input of the wire. Discrete band phase-matching $\Delta k_{nl} + \Delta k_l = 0$ is assumed for a moderate input pump power of 300 mW, consistent with the phase-matching calculation in Fig. 3b. However, optical loss mechanisms eventually result in a phase-mismatch, particularly as the pump is attenuated and/or depleted along the length of the wire. This accumulated phase-mismatch results in an optimal nanophotonic wire length, at which a maximum value of spectral translation efficiency is reached. As this optimal length depends strongly upon the linear propagation loss, wires with propagation losses of 0.6 dB/cm, 0.3 dB/cm, and 0.1 dB/cm are simulated, producing optimal lengths of 5 cm, 8.5 cm, and 20 cm, respectively. We note that air-clad silicon nanophotonic wires having significantly reduced propagation losses of 0.6 dB/cm across the 2000-2500 nm wavelength range have recently been demonstrated[14]. By substituting the air cladding for silicon dioxide as illustrated in Supplementary Fig. 4a, it is reasonable to assume that the reduced refractive index contrast could lead to even lower propagation losses in the range of 0.1-0.3 dB/cm.

Finally, the spectral translation efficiency and mid-IR signal gain are computed as a function of the continuous-wave pump power at 2200 nm, for each of the three optimal silicon nanophotonic wire lengths. As expected, Supplementary Fig. 5 illustrates that the translation efficiency and signal gain grow more rapidly as the wire propagation loss decreases. At a moderate pump power of 300 mW, Supplementary Fig. 5a shows that the 5 cm long wire with 0.6 dB/cm loss reaches signal transparency, while the translation efficiency is significant at -0.6 dB. For further reduced propagation losses of 0.3 dB/cm and 0.1 dB/cm, Supplementary Figs. 5b and 5c illustrate that a pump power of 300 mW produces a signal gain (translation efficiency) of 4.5 dB (5.4 dB) and 21 dB (22 dB), respectively. Moreover, a spectral translation efficiency of 0 dB is obtained with pump powers of 320 mW, 170 mW, and 65 mW for the 0.6 dB/cm, 0.3 dB/cm, and 0.1 dB/cm silicon wires, respectively. Therefore, these simulations indicate that dispersion-engineered silicon nanophotonic wires can facilitate highly efficient spectral translation of mid-IR signals for un-cooled telecom band detection, using continuous-wave pumping with reasonable values of on-chip pump power.

**Supplementary Figures:**

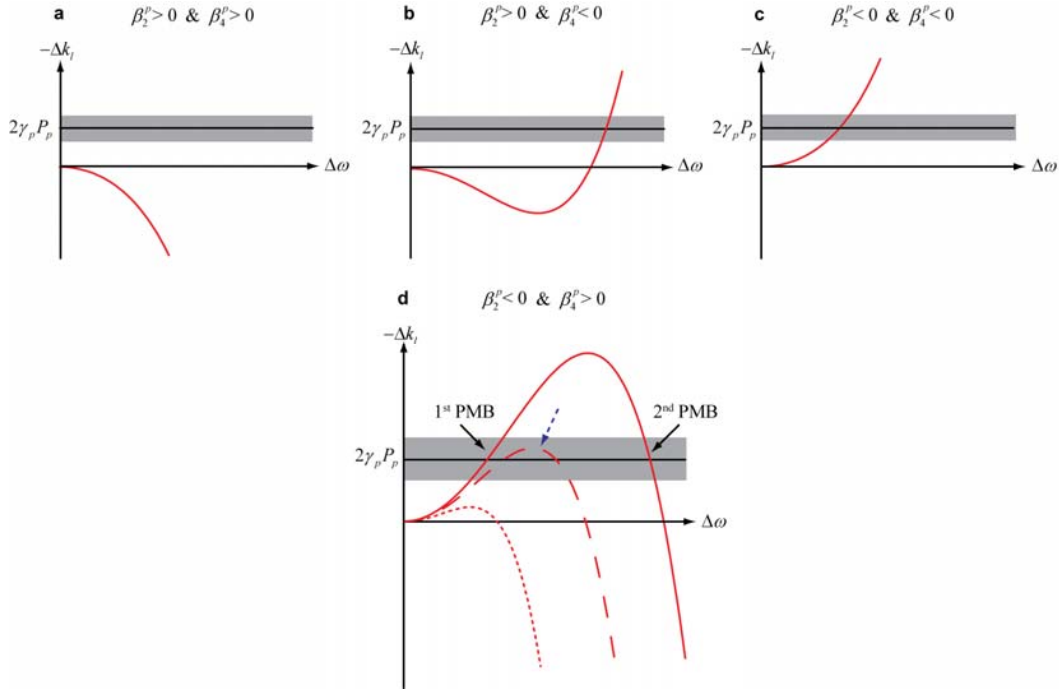

**Supplementary Figure 1 | Generalized illustration of graphical solutions to the phase-matching condition expressed in Eq. (2), for various signs of $\beta_2$ and $\beta_4$. a**, $\beta_2^p > 0$ and $\beta_4^p > 0$. **b**, $\beta_2^p > 0$ and $\beta_4^p < 0$. **c**, $\beta_2^p < 0$ and $\beta_4^p < 0$. **d**, $\beta_2^p < 0$ and $\beta_4^p > 0$. Solutions exist where the red curves, representing the frequency-detuning dependent linear phase-mismatch $-\Delta k_l$, intersect with the pump power-dependent nonlinear phase-mismatch term $2\gamma_p P_p$. The grey shaded region is intended to depict the phase detuning tolerance $\pm 2\pi/L$ due to the finite length $L$ of the silicon nanophotonic wire, within which efficient FWM can be achieved. For clarity, only the solutions on the anti-Stokes ($\Delta\omega > 0$) side of the pump have been shown.

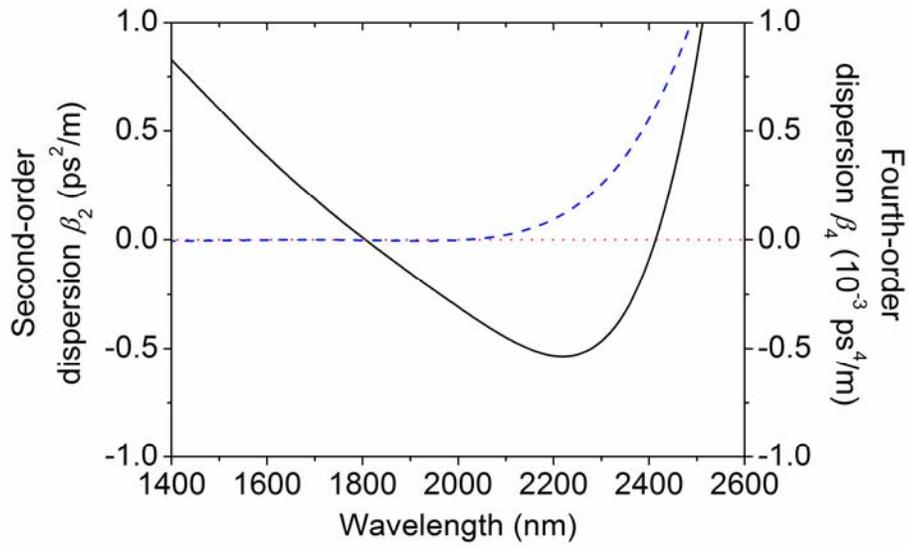

**Supplementary Figure 2 | Wavelength-dependent dispersion coefficients for the fundamental quasi-TE mode of the silicon nanophotonic wire spectral translation device depicted in Fig. 1.** Numerically calculated values for the $2^{nd}$-order dispersion $\beta_2$ and the $4^{th}$-order dispersion $\beta_4$ are illustrated by the solid black and dashed blue curves, respectively.

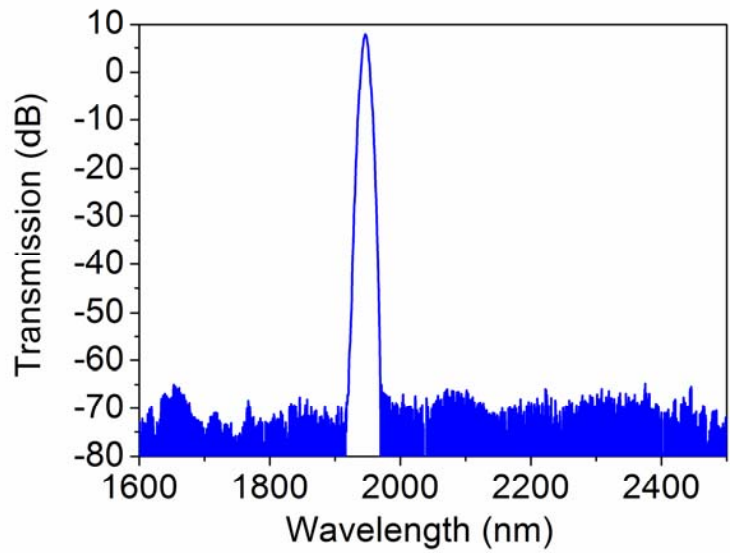

**Supplementary Figure 3 | Spectrum of the λ = 1946 nm pump pulse-train.** This spectrum illustrates that the pump pulse has an optical signal-to-noise ratio > 70 dB over the wavelength range from 1600 nm to 2500 nm.

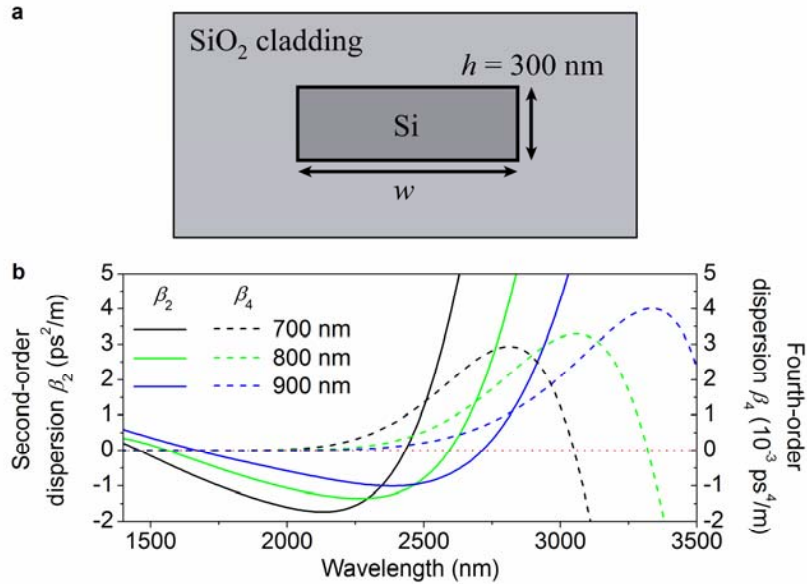

**Supplementary Figure 4 | Silicon nanophotonic wire designs optimized for translating λ = 3000-3550 nm mid-IR signals to the L-band. a**, Cross-section schematic of the silicon dioxide-clad silicon wire design. **b**, Simulated waveguide dispersion of the fundamental quasi-TE mode of wires having height $h$ = 300 nm and widths of $w$ = 700 nm (black), 800 nm (green), and 900 nm (blue) respectively. Solid curves represent second-order dispersion $\beta_2$, dashed curves depict fourth-order dispersion $\beta_4$. In comparison with the experimentally demonstrated silicon nanophotonic wire device shown in Fig. 1, this improved design possesses larger values of $|\beta_2|$ and smaller values of $|\beta_4|$ in the vicinity of silicon's TPA threshold at 2200 nm, facilitating spectral translation and amplification across more than a full octave in optical frequency. However, using the conventional SOI platform, it may be difficult to realize low-loss silicon nanophotonic wire mid-IR components operating at even longer wavelengths, due to the strong molecular absorption within the $SiO_2$ cladding at wavelengths longer than ~3600 nm[15, 16]. While the silicon wire core remains highly transmitting at wavelengths up to approximately 8000 nm[15, 17], the cladding and/or core materials could be replaced with materials such as germanium[15, 17],

sapphire[18, 19], silicon nitride[15, 20], or air[21]. Various combinations of these materials could be employed to address applications within the 3600-8000 nm spectral region.

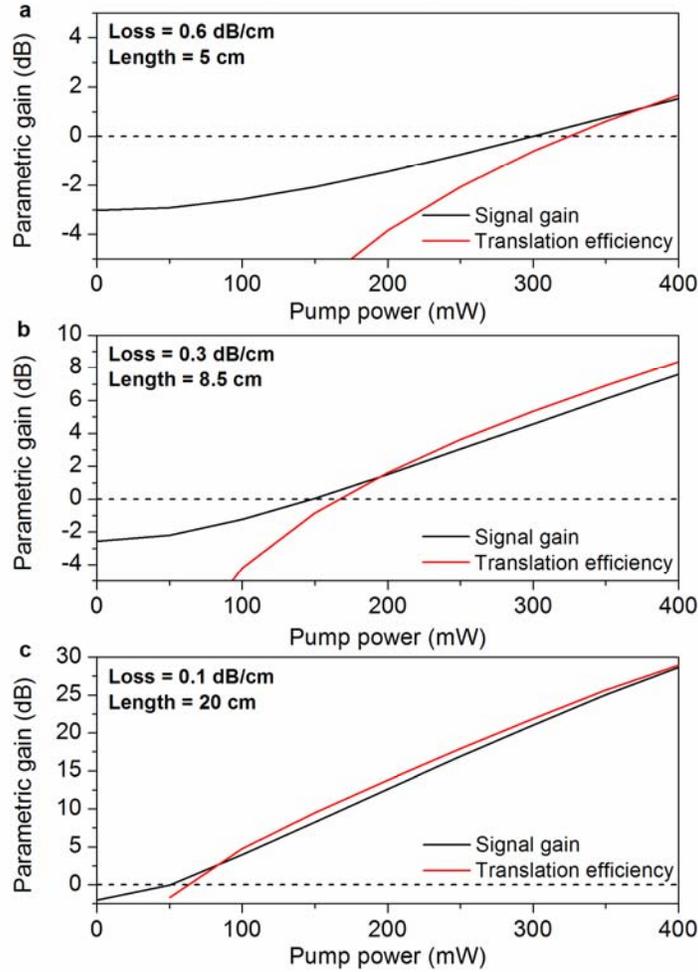

**Supplementary Figure 5 | Simulated parametric spectral translation efficiency and signal gain versus continuous-wave pump power, for an oxide-clad silicon nanophotonic wire with $w$ = 900 nm and $h$ = 300 nm.** Using a pump wavelength of 2200 nm, a mid-IR input signal at 3550 nm is spectrally translated to an L-band idler at 1590 nm. Silicon wires with linear propagation losses of **a**, 0.6 dB/cm, **b**, 0.3 dB/cm, and **c**, 0.1 dB/cm are considered, which have optimal lengths of 5 cm, 8.5 cm, and 20 cm respectively for a pump power of 300 mW. The pump power required to reach a translation efficiency of 0 dB (dashed black line) is 320 mW, 170 mW, and 65 mW for the 0.6 dB/cm, 0.3 dB/cm, and 0.1 dB/cm silicon wires, respectively.

**Supplementary Notes:**